\documentclass[twocolumn,aps,superscriptaddress,showpacs]{revtex4}
\usepackage{amsmath,bm}
\usepackage{graphicx}

\begin{document}

\title{Evolution of the symmetry energy of hot neutron-rich matter formed in
heavy-ion reactions}
\author{Bao-An Li}
\affiliation{Department of Physics, Texas A\&M University-Commerce, Commerce, TX 75429,
and Department of Chemistry and Physics, P.O. Box 419, Arkansas State
University, State University, AR 72467-0419, USA}
\author{Lie-Wen Chen}
\affiliation{Institute of Theoretical Physics, Shanghai Jiao Tong University, Shanghai
200240, China}
\affiliation{Center of Theoretical Nuclear Physics, National Laboratory of Heavy-Ion
Accelerator, Lanzhou, 730000, China}

\begin{abstract}
It is shown that the experimentally observed decrease of the
nuclear symmetry energy with the increasing centrality or the
excitation energy in isotopic scaling analyses of heavy-ion
reactions can be well understood analytically within a degenerate
Fermi gas model. The evolution of the symmetry energy is found to
be mainly due to the variation in the freeze-out density rather
than temperature. The isoscaling analyses are useful for probing
the interaction part of the nuclear symmetry energy, provided that
both the freeze-out temperature and density of the fragments can
be inferred simultaneously from the experiments.
\end{abstract}

\maketitle

\section{Introduction}

Information about the symmetry energy of hot neutron-rich matter is
important for understanding the dynamical evolution of massive stars and the
supernova explosion mechanisms, while the symmetry energy at zero
temperature is important for determining properties of neutron stars at $%
\beta $-equilibrium. In particular, the electron capture rate on nuclei
and/or free protons in pre-supernova explosions is especially sensitive to
the symmetry energy at finite temperatures. The electron captures drive the
collapsing core towards more neutron-rich matter. They affect not only the
electron degenerate pressure working against the gravity but also the
neutrino fluxes carrying away energy from the core\cite%
{bethe79,lat01,heger01,steiner05}. The larger the symmetry energy, the more
difficult it is for the electron captures to happen. Heavy-ion reactions are
a unique means to produce in terrestrial laboratories the hot neutron-rich
matter similar to those existing in many astrophysical situations. The
possibility of extracting useful information about the symmetry energy from
heavy-ion reactions has stimulated much interest in the nuclear physics
community \cite{ireview98,ibook01,ditoro}. Especially, recent analyses of
the isospin diffusion data in heavy-ion reactions \cite%
{betty04,chen05,lichen05} and the size of neutron-skin in $^{208}$Pb \cite%
{steiner,jorge05,chenkl} have already put a stringent constraint on the
symmetry energy of cold neutron-rich matter at sub-normal densities. This
has led to a significantly more refined constraint on the mass-radius
correlation of neutron stars \cite{listeiner} including the fastest pulsar
discovered very recently \cite{science}. On the other hand, the temperature
dependence of the symmetry energy for hot neutron-rich matter has received
so far little theoretical attention.

Among the phenomena/observables identified as potentially useful probes of
the nuclear symmetry energy, the isoscaling coefficients of fragments from
heavy-ion reactions \cite{betty01} have been most extensively studied, see
e.g., \cite{maria} for a recent review. Very interestingly, it was found
recently that the extracted symmetry energy from the isoscaling analyses
decreases significantly from its standard value of about $25$ MeV at normal
nuclear matter density $\rho _{0}$ to much smaller values with the
increasing excitation energy or centrality in heavy-ion reactions at both
Fermi \cite{shetty,sjy} and relativistic energies \cite{indra,henz,wolfgang}%
. Moreover, an increasing temperature of the fragmenting system was found to
accompany the decreasing symmetry energy in these reactions. However, the
fundamental origin of this apparent evolution of the symmetry energy is
still not clear and it is particularly important and interesting to
understand to what degree the evolution is due to the density and/or the
temperature dependence of the symmetry energy.

In this work, it is shown that the experimentally observed evolution of the
symmetry energy can be well understood within a degenerate Fermi gas model
at finite temperatures. Furthermore, it is found that the evolution of the
symmetry energy is mainly due to the variation in the freeze-out density
rather than temperature when the fragments are emitted in the reactions
carried out under different conditions.

\section{Nuclear Symmetry Energy at Finite Temperature}

The Equation of State (EOS) of hot neutron-rich matter at a temperature $T$
and an isospin asymmetry $\delta \equiv (\rho _{n}-\rho _{p})/(\rho
_{n}+\rho _{p})$ can be written as \cite{chen01,zuo03}
\begin{equation}
E(\rho ,T,\delta )=E(\rho ,T,\delta =0)+E_{sym}(\rho ,T)\delta ^{2}+\mathcal{%
O}(\delta ^{4}).  \label{eos}
\end{equation}%
The temperature and density dependent symmetry energy $E_{sym}(\rho ,T)$ for
hot neutron-rich matter can thus be extracted from $E_{sym}(\rho ,T)\approx
E(\rho ,T,\delta =1)-E(\rho ,T,\delta =0)$. The symmetry energy $%
E_{sym}(\rho ,T)$ is the energy cost to convert all protons in symmetry
matter to neutrons at the fixed temperature $T$ and density $\rho $. For
finite nuclei at temperatures below about $3$ MeV, the shell structure and
pairing as well as vibrations of nuclear surfaces are important and the
symmetry energy was predicted to increase slightly with the increasing
temperature \cite{don94,dean}. Interestingly enough, an increase by only
about $8\%$ in the symmetry energy in the range of $T$ from $0$ to $1$ MeV
was found to affect appreciably the physics of stellar collapse, especially
the neutralization processes\cite{don94}. At higher temperatures, one
expects the symmetry energy to decrease as the Pauli blocking becomes less
important when the nucleon Fermi surfaces become more diffused at
increasingly higher temperatures \cite{chen01,zuo03}. In this work, we use
the thermal model of Mekjian, Lee and Zamick (MLZ) \cite{mlz}. While all of
our results have also been concurrently verified numerically by using the
finite temperature Hartree-Fock (HF) approach using both Skyrme and Gogny
forces \cite{chen06}, here we utilize the MLZ approach because of its
analytically transparent properties. The results obtained within the MLZ
approach are sufficient for the purposes of this work. Our studies based on
the finite temperature HF calculations will be reported elsewhere \cite%
{chen06}. The degenerate Fermi gas limit of the MLZ thermal model is
appropriate for us to understand quantitatively the experimentally observed
evolution of the nuclear symmetry energy. The symmetry energy $E_{sym}(\rho
,T)$ has a kinetic contribution and an interaction part. In the MLZ model
with Skyrme interactions, the interaction part is temperature independent,
i.e.,
\begin{equation}
E_{sym}(\rho ,T)=E_{sym}^{kin}(\rho ,T)+E_{sym}^{int}(\rho ).
\end{equation}
At low temperatures, it is known that all mean field quantities are
essentially temperature independent \cite{don94}. With even
momentum-dependent Gogny forces, at temperatures relevant for fragment
formation in heavy-ion reactions, our HF calculations indicate that the
interaction part $E_{sym}^{int}$ is only slightly $T$-dependent. For
temperatures much less than the Fermi energy, $T\ll E_{Fermi}\approx 36(\rho
/\rho _{0})^{2/3}$, the kinetic energy per nucleon of a near-degenerate
two-component Fermi gas is \cite{mlz}
\begin{equation}
E_{kin}(\rho ,T,\delta )=21u^{2/3}(1+\frac{5}{9}\delta ^{2})+\frac{\pi ^{2}}{%
140}\frac{T^{2}}{u^{2/3}}(1-\frac{1}{9}\delta ^{2}),
\end{equation}%
where $u=\rho /\rho _{0}$ is the reduced density. The kinetic part of the
symmetry energy is thus
\begin{equation}
E_{sym}^{kin}(\rho ,T)=\frac{35}{3}u^{2/3}-\frac{\pi ^{2}}{1260}\frac{T^{2}}{%
u^{2/3}}.
\end{equation}%
It is interesting to note that the $E_{sym}^{kin}(\rho ,T)$ decreases with $%
-T^{2}$ with a rate depending on the density. On the other hand, for $T\gg
E_{Fermi}$ the system becomes a non-degenerate classical gas with a kinetic
symmetry energy of \cite{mlz}
\begin{equation}  \label{sym2}
E_{sym}^{kin}(\rho ,T)=\frac{3}{2}T[0.177\frac{\lambda ^{3}}{4}\rho -0.0033(%
\frac{\lambda ^{3}}{4}\rho )^{2}],
\end{equation}%
where $\lambda =\sqrt{2\pi \hbar ^{2}/mT}$ is the thermal wavelength of
nucleons with an average mass $m$. It is seen that the symmetry energy
decreases approximately according to $E_{sym}(\rho ,T)\propto \rho /\sqrt{T}$
for $\lambda ^{3}\rho \ll 1$. The genuine feature of a decreasing symmetry
energy associated with an increasing temperature at both the degenerate and
non-degenerate Fermi gas limits is consistent with predictions of the
microscopic and/or phenomenological many-body theories \cite%
{chen01,zuo03,horowitz,rou,chen06}.

At the typical freeze-out temperatures and densities for the emission of
intermediate mass fragments in heavy-ion reactions, the degenerate Fermi gas
limit is appropriate except at extremely low densities where clustering
effects may become important \cite{horowitz,joe}. To evaluate the relative
importance of the $T$-dependent part $\bigtriangleup E^{T}\equiv \pi
^{2}T^{2}/(1260u^{2/3})$ with respect to the total symmetry energy for cold
neutron-rich matter, we show in Fig. 1 the ratio $\bigtriangleup
E^{T}/E_{sym}(\rho ,T=0)$. The analyses of the isospin diffusion data from
NSCL/MSU \cite{betty04,chen05,lichen05} and the study on the size of
neutron-skin in $^{208}$Pb \cite{steiner,jorge05,chenkl} have recently
consistently constrained the symmetry energy of cold matter to be around $%
32(\rho /\rho _{0})^{0.7}\leq E_{sym}(\rho ,T=0)\leq 32(\rho /\rho
_{0})^{1.1}$ at sub-normal densities. Using the above two limits for the $%
E_{sym}(\rho ,T=0)$ at a typical freeze-out temperature of $T=5$ MeV for the
fragment emission in heavy-ion reactions, it is seen that the ratio $%
\bigtriangleup E^{T}/E_{sym}(\rho ,T=0)$ increases quickly with decreasing
density. The $T$-dependent part of the symmetry energy becomes increasingly
more appreciable, e.g., up to about $35\%$ at $\rho =0.1\rho _{0}$ for $%
E_{sym}(\rho ,T=0)=32(\rho /\rho _{0})^{1.1}$. Moreover, since the ratio
increases quadratically with $T$, the effect will be much larger at higher
temperatures. This result also shows the magnitude, especially at low
densities and/or high temperatures, of some artificial effects that would be
introduced should one attribute the entire evolution of the symmetry energy
to its density dependence while neglecting its intrinsic temperature
dependence.
\begin{figure}[tbh]
\includegraphics[width=5cm,height=7.6cm,angle=-90]{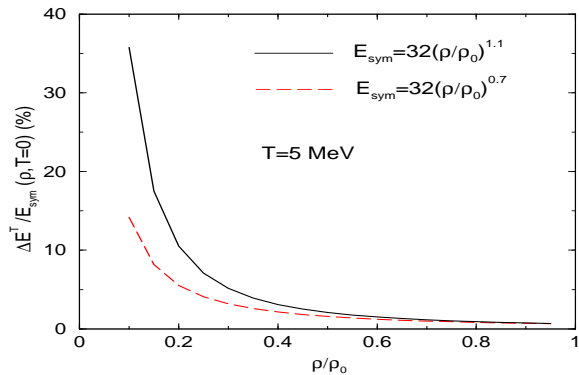}
\caption{{\protect\small (Color online) The relative importance of
the temperature-dependent part of the nuclear symmetry energy.}}
\label{lifig1}
\end{figure}

Corresponding to the above two limits for the $E_{sym}(\rho
,T=0)$, the interaction part of the symmetry energy
$E_{sym}^{int}(\rho )$ is constrained between $E_{sym}^{int}(\rho
)=130u-111.4u^{1.1}$ (labeled as $x=0
$) and $E_{sym}^{int}(\rho )=3.7u+14.9u^{1.6}$ (labeled as $x=-1$) \cite%
{chen05,lichen05}. With this $E_{sym}^{int}(\rho )$ and Eq. (4), we can now
examine the evolution of the total symmetry energy (Eq. (2)) as functions of
density and temperature. Shown in Fig.\ 2 is the evolution of the normalized
symmetry energy $[E_{sym}(\rho ,T)-E_{sym}(\rho _{0},0)]/E_{sym}(\rho _{0},0)
$ as a function of temperature at sub-saturation densities. It is seen that
in the temperature range considered here, the symmetry energy does not
change much with temperature at a given density. It is interesting to
mention that using the same Gogny interactions corresponding to $x=0$ and $%
x=-1$ \cite{das03} in the finite temperature HF approach, all of the major
results discussed above are qualitatively reproduced \cite{chen06}. The
major physical features given by the MLZ thermal model are thus rather
general.

\section{Isoscaling in Heavy-Ion Collisions}

It has been observed in many types of reactions that the ratio $R_{21}(N,Z)$
of yields of a fragment with proton number $Z$ and neutron number $N$ from
two reactions reaching about the same temperature $T$ satisfies an
exponential relationship $R_{21}(N,Z)\propto exp(\alpha N)$\cite%
{betty01,shetty,sjy,indra,henz,tsang01,botvina,ono,dorso,ma}. In several
statistical and dynamical models under some assumptions\cite%
{tsang01,botvina,ono}, it has been shown that the scaling coefficient $%
\alpha $ is related to the symmetry energy $C_{sym}(\rho ,T)$ via
\begin{equation}
\alpha =\frac{4C_{sym}(\rho ,T)}{T}\bigtriangleup \lbrack (Z/A)^{2}],
\label{scaling}
\end{equation}%
where $\bigtriangleup \lbrack (Z/A)^{2}]\equiv
(Z_{1}/A_{1})^{2}-(Z_{2}/A_{2})^{2}$ is the difference between the $(Z/A)^{2}
$ values of the two fragmenting sources created in the two reactions.

Before we proceed, a few comments and discussions regarding the validity of
Eq. (\ref{scaling}) and the physical meaning of $C_{sym}$ are in order.
First of all, we notice that Eq. (\ref{scaling}) is an approximation in
equilibrium models and an empirical assumption in dynamical models where isoscaling is
observed in generated events. Because of the different assumptions used in the various
derivations, the validity of this equation is still disputable as to whether
and when the $C_{sym}$ is actually the symmetry energy or the symmetry free
energy. Moreover, the physical interpretation of the $C_{sym}(\rho ,T)$ is
also not clear, sometimes even contradictory, in the literature.
The main issue is whether the $C_{sym}$ measures the symmetry energy of the
fragmentating source or that of the fragments formed at freeze-out.
This ambiguity is also due to the fact that the derivation of Eq. (\ref{scaling})
is not unique. In particular, within the grand canonical statistical model for
multifragmentation\cite{tsang01,botvina} the $C_{sym}$ refers to the symmetry energy of
primary fragments. While within the sequential Weisskopf model in the grand canonical limit\cite{tsang01}
it refers to the symmetry energy of the emission source.

Based on the grand canonical model for multifragmentation, some experts
take the view that the $C_{sym}$ is the symmetry energy of the fragments at the
freeze-out. Moreover, they use the finite size effects, such as the surface symmetry
energy and its temperature dependence, to explain the observed smaller value of
$C_{sym}$ compared to the symmetry energy of about 30 MeV for cold nuclear matter
at the saturation density. However, from the very nature of the
isoscaling phenomenon itself that isotopes/isotones having very different
masses (sizes) fall on the same curve described by a single scaling
coefficient, it is hard to believe that the finite-size effects have any
influence on the $C_{sym}$ at all. In another word, unless the finite size effects on
both the $C_{sym}$ and the temperature $T$, if there is any at all, are completely cancelled
out, the isoscaling phenomenon should not have been observed in multifragmentation in the
first place. Indeed, in the AMD analyses of isoscaling in multifragmentation,
it was found that ``{\it the extracted symmetry energy shows almost no surface effect in it,
which suggests that the properties of infinite nuclear matter can be directly obtained
from the information of fragmentation}''\cite{ono04}.

Within the grand canonical model for multifragmentation, another
possible reason for the extracted small value of $C_{sym}$
is that fragments themselves at freeze-out are dilute.
This picture, however, seems to contradict the
basic Fisher hypothesis that correlations inside a dilute medium
are exhausted by clusterization. One possible explanation put
forward was that primary fragments formed in heavy-ion reactions
are hot and thus also expanded \cite{shetty,botvina}. However,
this explanation is insufficient to explain the much smaller
value of $C_{sym}$ observed in central collisions. Moreover, the isoscaling
phenomenon is actually observed experimentally for cold fragments.
The sequential decay of hot primary fragments may not affect much
the isoscaling coefficient. We notice that this is still a matter of hot debate
depending on the model calculations \cite{ono05,shetty06}.
Therefore, with this view the small value of $C_{sym}$ for cold
fragments extracted in the isoscaling experiments would indeed
indicate that the fragments have dilute internal density. This
would require a deep reconsideration of the statistical models
from which Eq. (\ref{scaling}) was derived.

To this end, it is necessary to repeat here some remarks made by
one of us at the WCI3 meeting\cite{wci3}. While the
Eq.(\ref{scaling}) is a good approximation within the grand
canonical statistical model for multifragmentation, the isoscaling
coefficient $\alpha$ is sensitive to the density dependence of the
symmetry energy {\it not} because of the $C_{sym}$ which should be
the symmetry energy of normal nuclear matter, {\it but rather}
because of the $Z/A$ ratios of the fragmentating sources through
the dynamical isospin fractionation\cite{li00} in the early stage
of the reaction. This was also pointed out recently in
ref.\cite{ganil}. Unfortunately, since the $Z/A$ ratio of the
effective fragmentating source, if exists at all, is not directly
accessible experimentally, simplified assumptions are normally
made in the data analyses within both statistical and dynamical
models. The efforts are normally misplaced on extracting the
$C_{sym}$ as if it is the one depending on the density and
temperature. Consequently, the extracted $C_{sym}$ may thus
contain some information about the density dependence of the
symmetry energy that is actually carried by the $Z/A$ ratios of
the fragmentating sources.

On the other hand, within the sequential Weisskopf model in the grand canonical limit\cite{betty01},
all quantities in Eq. (\ref{scaling}) refer to the emission source. In particular,
the $C_{sym}$ itself in Eq. (\ref{scaling}) reflects the bulk symmetry energy of the low density
fragmentating source. Besides, because of the way by which the data is analyzed as we
mentioned above, the experimentally extracted $C_{sym}$ also contains information
about the symmetry energy through the $Z/A$ ratios of the fragmentatig sources.
Here we thus use broadly a working assumption that the $C_{sym}$ reflects the
symmetry energy of bulk nuclear matter. We notice that this assumption
can be justified only in the sequential weisskopf model.
Nevertheless, it is interesting and reassuring to
note that this assumption is consistent with the statement in
Ref. \cite{ono05} that \textit{the $C_{sym}$ is the symmetry energy
of uniform nuclear matter at a reduced density}. Within this picture it
is natural for the $C_{sym}$ to have values smaller than the symmetry energy of
normal nuclear matter. One would thus have no difficulty with the Fisher hypothesis for fragment formation.
At freeze-out, fragments are formed at their normal density in a large volume.
Only the average density in the freeze-out volume is small.

We would also like to stress here that the $C_{sym}(\rho ,T)$ extracted from studying the
isoscaling coefficient $\alpha $ is the total symmetry energy at the finite
temperature $T$. It should be distinguished from the symmetry energy at zero
temperature extracted from transport model analyses of dynamical
observables, such as the isospin diffusion, or that from studying the
neutron-skins of heavy nuclei. In transport models, the nucleon mean field
or effective interaction for cold nuclear matter is used as an input. The
zero temperature symmetry energy as a function of density can thus be
constructed analytically from the particular mean fields or effective
interactions used in the calculations. Special cares should thus be taken
when comparing the density functions for nuclear symmetry energy extracted
from different approaches and/or results from the same approach but for
reactions at different temperatures, especially at low densities.
\begin{figure}[tbh]
\includegraphics[width=7.5cm,height=8cm,angle=-90]{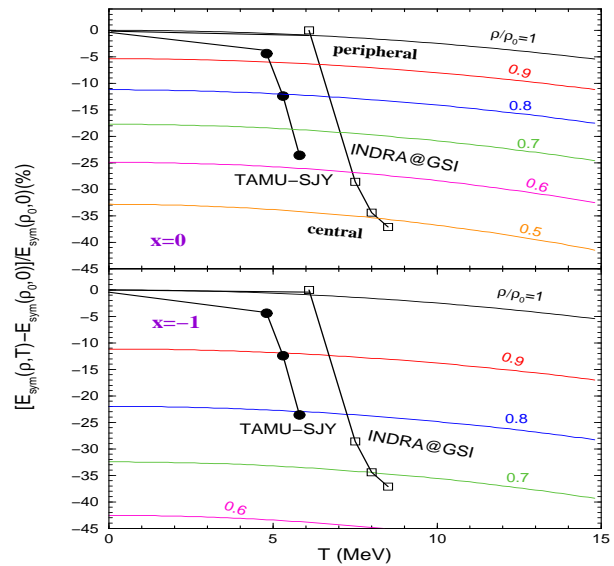}
\caption{{\protect\small (Color online) The relative evolution of
symmetry energy as functions of temperature and density with two
different interaction-parts of the symmetry energy (see text).}}
\label{lifig2}
\end{figure}

Within our working assumption that the extracted $C_{sym}$ from isoscaling analyses is the
symmetry energy of the fragmenting source, we can then compare the experimental
$C_{sym}$ with our calculations of the symmetry energy for uniform nuclear matter.
In Fig.\ 2, the experimental data taken from the SJY group at Texas A\&M
University (filled circles) \cite{shetty,sjy} and the INDRA-ALADIN
collaboration at GSI (open squares) \cite{indra,wolfgang} are compared with
the calculations. In the TAMU experiments, combinations of projectile-like
fragments from peripheral to semiperipheral collisions of $25$ MeV/nucleon $%
^{86}$Kr and $^{64}$Ni beams on several neutron-rich targets were used in
the isoscaling analyses. The symmetry energy was found to decrease quickly
from about $25$ MeV to $19$ MeV as the temperature increases from about $4.8$
MeV to $5.8$ MeV \cite{sjy}. The INDRA@GSI data were obtained from the
fragmentation of target-like residues following collisions of $^{12}$C on $%
^{112,124}$Sn targets at a beam energy of $300$ MeV/nucleon. The INDRA@GSI
data indicate that the symmetry energy decreases from about $26$ MeV to $16$
MeV as the temperature increases from about $6$ MeV to $9$ MeV when the
reaction goes from peripheral to central collisions \cite{indra}.

It is very interesting to compare the calculations using both the $x=0$
(upper window) and the $x=-1$ (lower window) interactions with the
experimental data. The comparison then allows us to estimate the required
density of the fragment emitting source. At the respectively low
temperatures reached in the peripheral reactions at TAMU and GSI, the
calculated results indicate that the fragments are emitted from sources at
densities only slightly below $\rho _{0}$. In the peripheral reactions,
either the projectile-like or target-like residue is only slightly excited
with little expansion. While at the higher temperatures reached in the more
central reactions, the fragments are emitted from significantly diluted
sources with densities depending on the interaction used. This picture is
consistent with dynamical model calculations of nuclear multifragmentations
in the energy range considered. More interestingly, it is seen that the
experimentally observed evolution of the symmetry energy is mainly due to
the change in density rather than temperature. Around the typical freeze-out
temperatures reached in both the TAMU and the GSI experiments, the evolution
of the symmetry energy due to the change in temperature at a given density
is rather small. It implies that the underlying origin of the observed
decrease of the symmetry energy with the apparently increasing temperature
or centrality in these experiments is actually due to the accompanying
decrease in freeze-out density. Comparing the two sets of data, it is seen
that they actually parallel with each other in the common density range.
Since the evolution of the symmetry energy is essentially independent of the
temperature for the experiments considered, the two sets of date thus
indicate the same density-dependence of the symmetry energy as one expects.
On the other hand, within the view that the extracted $C_{sym}$ reflects the
symmetry energy of the fragments at freeze-out, the two data sets are incompatible.

From the above discussions, we can see that the evolution of the symmetry
energy can be useful in exploring the density dependence of the interaction
part of the nuclear symmetry energy. The latter is most uncertain but very
important for many interesting questions in astrophysics. With the
interaction labeled $x=0$, the hottest point requires an average freeze-out
density of about $0.62\rho _{0}$ and $0.49\rho _{0}$ for the TAMU-SJY and
the INDRA@GSI data, respectively. While using the interaction labeled $x=-1$%
, the corresponding average freeze-out density is about $0.8\rho _{0}$ and $%
0.68\rho _{0}$, respectively. Therefore, the required freeze-out density
depends strongly on the interaction part of the symmetry energy. The effect
of using interactions of $x=0$ and $x=-1$ is about $30\%$ and $40\%$ for the
TAMU and the INDRA@GSI experiment, respectively. Of course, for the purpose
of extracting the interaction part of the symmetry energy from the
isoscaling analyses, it is necessary to know not only the freeze-out
temperature but also the density when the fragments are emitted from the
reactions. Fortunately, inferring both the freeze-out temperature and
density in isoscaling analyses has been shown feasible very recently by the
TAMU-JBN group \cite{joe}, albeit largely based on model calculations. Since
an independent determination of the freeze-out density will be very useful,
observables known to be sensitive to the freeze-out density, such as the
fragment correlation functions\cite{boal,bauer} or source functions from
imaging techniques\cite{pawel,david}, may be explored together with the
isoscaling analyses.

\section{Summary}

In summary, within the degenerate Fermi gas model of Mekjian, Lee and Zamick
it is shown that the experimentally observed evolution of the symmetry
energy can be well understood. Furthermore, the evolution is found to be
mainly due to the variation in the freeze-out density rather than
temperature in the reactions carried out under different conditions. The
isoscaling analyses are thus useful for probing the interaction part of the
nuclear symmetry energy, provided that both the freeze-out temperature and
density can also be inferred.

\begin{acknowledgments}
This work was supported in part by the US National Science Foundation under
Grant No. PHYS-0354572 and PHYS-0456890, and the NASA-Arkansas Space Grants
Consortium award ASU15154 (BAL); and by the National Natural Science
Foundation of China under Grant No. 10575071, MOE of China under project
NCET-05-0392, and Shanghai Rising-Star Program under Grant No. 06QA14024
(LWC).
\end{acknowledgments}

\end{document}